\documentclass[preprint,aps,prd,showpacs,superscriptaddress,nofootinbib,tightenlines]{revtex4}

\usepackage{amsfonts}
\usepackage{multirow}
\usepackage{mathrsfs}
\usepackage{graphicx}
\usepackage{amsmath}
\usepackage{amssymb}
\usepackage{bm}
\usepackage{bbm}
\usepackage{color}
\usepackage{array}
\usepackage{diagbox}
\usepackage{float}
\usepackage{makecell}
\allowdisplaybreaks[4]

\def\OMIT#1{}

\def\hlinew#1{%
  \noalign{\ifnum0=`}\fi\hrule \@height #1 \futurelet
   \reserved@a\@xhline}
\usepackage{array}
\newcommand{\PreserveBackslash}[1]{\let\temp=\\#1\let\\=\temp}
\newcolumntype{C}[1]{>{\PreserveBackslash\centering}p{#1}}
\newcolumntype{R}[1]{>{\PreserveBackslash\raggedleft}p{#1}}
\newcolumntype{L}[1]{>{\PreserveBackslash\raggedright}p{#1}}

\newcommand{\nn}{\nonumber}

\newcommand{\beq}{\begin{equation}}
\newcommand{\eeq}{\end{equation}}
\newcommand{\bqa}{\begin{eqnarray}}
\newcommand{\eqa}{\end{eqnarray}}

\newcommand\fverb{\setbox\fverbbox=\hbox\bgroup\verb}
\newcommand\fverbdo{\egroup\medskip\noindent%
			\fbox{\unhbox\fverbbox}\ }
\newcommand\fverbit{\egroup\item[\fbox{\unhbox\fverbbox}]}
\newbox\fverbbox

\makeatletter

\newcommand{\Rmnum}[1]{\expandafter\@slowromancap\romannumeral #1@}
\makeatother

\begin{document}
%\preprint{}
%%%%%%%%%%%%%%%%%%%%%%%%%%%%%%%%%%%%%%%%%%%%%%%%%%%%%%%%%%%%%%%%%%%%%%%%%%%%%%
\title{\mbox{}\\[10pt]
Two-Loop QCD Corrections to C even Bottomonium Exclusive Decays to Double $J/\psi$}

\author{Yu-Dong Zhang~\footnote{ ydzhang@mails.ccnu.edu.cn}}
 \affiliation{Institute of Particle Physics and Key Laboratory of Quark and Lepton Physics (MOE),
 	Central China Normal University,Wuhan, Hubei 430079, China\vspace{0.2cm}}
 
 \author{Xiao-Wei Bai~\footnote{xiaoweibai22@163.com}}
 \affiliation{School of Physical Science and Technology, Southwest University, Chongqing 400700, China\vspace{0.2cm}}
 
\author{Feng Feng~\footnote{f.feng@outlook.com}}
\affiliation{Institute of High Energy Physics and Theoretical Physics Center for Science Facilities, Chinese Academy of Sciences, Beijing 100049, China\vspace{0.2cm}}
\affiliation{China University of Mining and Technology, Beijing 100083, China\vspace{0.2cm}}
 
 \author{Wen-Long Sang~\footnote{wlsang@swu.edu.cn}}
 \affiliation{School of Physical Science and Technology, Southwest University, Chongqing 400700, China\vspace{0.2cm}}
 
 \author{Ming-Zhen Zhou~\footnote{zhoumz@swu.edu.cn}}
 \affiliation{School of Physical Science and Technology, Southwest University, Chongqing 400700, China\vspace{0.2cm}}
 
%%%%%%%%%%%%%%%%%%%%%%%%%%%%%%%%%%%%%%%%%%%%%%%%%%%%%%%%%%%%%%%%%%%%%%%%%%%%%%
\date{\today}
%%%%%%%%%%%%%%%%%%%%%%%%%%%%%%%%%%%%%%%%%%%%%%%%%%%%%%%%%%%%%%%%%%%%%%%%%%%%%%
\begin{abstract}
In the framework of nonrelativistic QCD (NRQCD) factorization, we compute both the polarized and the unpolarized decay widths for the processes $\eta_b(\chi_{bJ})\to J/\psi J/\psi$, accurate up to next-to-next-to-leading-order (NNLO) in $\alpha_s$.  
For the first time, we confirm that the NRQCD factorization does hold at NNLO for the process involving triple quarkonia.
We find the radiative corrections are considerable. 
In particular for $\chi_{b2}$, both $\mathcal{O}(\alpha_s)$ and $\mathcal{O}(\alpha_s^2)$ corrections are sizable and negative,  and can significantly reduce the leading order prediction.
At NNLO, the branching fractions are $8.2\times 10^{-7}$, $6.2\times 10^{-6}$, $7.2\times 10^{-7}$ and $2.7\times 10^{-6}$ for $\eta_b$, $\chi_{b0}$,
$\chi_{b1}$ and $\chi_{b2}$ decay, respectively. 
Our theoretical predictions are consistent with the upper limits measured by the {\tt Belle} Collaboration. 
Moreover, we investigate the dependence of the theoretical predictions on
the ratio of the charm quark mass and the bottom quark mass, i.e., $r=m_c/m_b$. By fixing $m_b$ and varying $m_c$ from $1.25$ to $1.9$ GeV, we find the 
branching fraction can change a factor of $2$, $3$, and $6$ for $\eta_b$, $\chi_{b0}$, and $\chi_{b1}$, respectively.
Although the  branching fraction for $\chi_{b2}$ decreases with the increase of $r$ at leading order and next-to-leading order, it is almost independent on $r$ at NNLO.
In the phenomenological analysis, with the integrate luminosity $\mathcal{L}=100\,{\rm fb}^{-1}$, 
we expect about $(5-10)\times 10^3$ $\eta_b(\chi_{bJ})\to J/\psi J/\psi\to \ell \bar{\ell}\ell \bar{\ell}$  events produced at the {\tt LHC},
thus it might be hopeful to search for these processes. 
On the other hand,  there are less than $100$ $\eta_b(\chi_{bJ})\to J/\psi J/\psi$ signal events at the B factory, so it seems the experimental measurements on these channels are quite challenging based on current  dataset. 
Nevertheless, with the designed $50\, {\rm ab}^{-1}$ integrated luminosity at {\tt Belle} 2, 
the observation prospects of $\eta_b(\chi_{bJ})\to J/\psi J/\psi$
may be promising in the foreseeable future.

\end{abstract}
\maketitle

%---------------------------
\section{introduction}
%---------------------------
The exclusive decay of a bottomonium into double charmonia provides an excellent testing ground to explore the 
interplay between perturbative and nonperturbative aspects of the QCD.  These processes can be studied in the framework of nonrelativistic QCD (NRQCD)
factorization formalism~\cite{Bodwin:1994jh}, which offers a systematic way to separate the short-distance effects and long-distance effects.
The experimentalists have made many attempts to search for such processes.
Based on enormous $\Upsilon(1S)$ and $\Upsilon(2S)$ events,  the {\tt Belle} Collaboration has measured the branching fraction for $\Upsilon(1S)\to J/\psi\, \chi_{c1}$, and set the upper limits for the branching fractions of $\Upsilon(nS)\to J/\psi\, \eta_c$ and $\Upsilon(nS)\to J/\psi\, \chi_{c0,2}$~\cite{Belle:2014wam}.
Besides, the search for double charmonium decays of the P-wave spin-triplet bottomonium states was performed in~\cite{Shen:2012ei}.  Although
no significant $\chi_{bJ}$ signal is observed,  the upper limits for the branching fractions of $\chi_{bJ}\to J/\psi J/\psi$ were obtained~\cite{Shen:2012ei}. 

To date, the processes of a bottomonium exclusive decay into double charmonia have been extensively studied on the theoretical side.  For $\Upsilon$ decay,
the process of $\Upsilon\to J/\psi \eta_c$ was first studied  a long time ago by Jia~\cite{Jia:2007hy} within the framework of NRQCD. 
The rate of $\Upsilon\to J/\psi \chi_{cJ}$ was first computed in Ref.~\cite{Xu:2012uh}.
The relativistic corrections and radiative corrections to these processes were separately considered in Ref.~\cite{Sang:2015owa} and Ref.~\cite{Zhang:2022nuf}.
For C even bottomonium decay,
due to kinematic constraint, the amplitude of $\eta_b\to J/\psi J/\psi$ disappears at lowest order in heavy quark velocity and $\alpha_s$. 
The relativistic corrections to the rate of $\eta_b\to J/\psi J/\psi$ were first calculated in~\cite{Jia:2006rx}.
In the same year, the next-to-leading order (NLO) radiative corrections were carried out~\cite{Gong:2008ue}.
The process was restudied based on light-cone (LC) approach~\cite{Braguta:2009xu}.
In 2010, Sun {\it et al.} recomputed the NLO corrections to the rate of $\eta_b\to J/\psi J/\psi$ in NRQCD~\cite{Sun:2010qx}. 
Moreover, they also calculated the higher twist effects in the LC formalism.  
For $\chi_{bJ}\to J/\psi J/\psi$, the decay rate was first computed by Braguta {\it et al.} both in NRQCD and LC formalism~\cite{Braguta:2010zz}.
Later, the relativistic corrections in charm quark velocity were carried out in Refs.~\cite{Zhang:2011ng,Sang:2011fw}, and the NLO radiative corrections were 
worked out in Ref.~\cite{Chen:2014lqa}.

In recent years, technological advances have made it possible to calculate the higher-order QCD corrections to quarkonium production and decay, in particular for
the processes involving multiple quarkonia.
The two-loop radiative corrections to the cross section of $e^+e^-\to J/\psi \eta_c$ at the B factory were computed in Refs.~\cite{Feng:2019zmt,Huang:2022dfw}.
In the last year, the two-loop corrections to $e^+e^-\to J/\psi \chi_{cJ}$ were obtained in Ref.~\cite{Sang:2022kub}.
With all the available radiative corrections lumped together,  
the theoretical results on the production cross sections of $J/\psi+\eta_c(\chi_{c0})$ agree with the experimental measurements, 
notwithstanding large uncertainties. 
Very recently, the very challenging two-loop radiative corrections to $e^+e^-\to J/\psi J/\psi$ were carried out~\cite{Sang:2023nvt}.
With the measured $J/\psi$ decay constant as input, which amounts to resumming a specific class of
radiative and relativistic corrections to all orders,  
the perturbative corrections exhibit a decent convergence behavior.
The two-loop corrections to the decay width of $\Upsilon\to \eta_c(\chi_{cJ})\gamma$ were evaluated in 2021~\cite{Zhang:2021ted}.
In addition, the two-loop corrections to $\Upsilon\to J/\psi\eta_c(\chi_{cJ})$ were worked out last year~\cite{Zhang:2022nuf}, 
where the QCD corrections notably mitigate the renormalization scale dependence of the decay widths, and the theoretical predictions on the branching fraction of $\Upsilon\to J/\psi\chi_{c1}$ is well consistent with the {\tt Belle} measurement~\cite{Belle:2014wam}. 
Inspired by the success of the NRQCD, we calculate, in the current work, the $\mathcal{O}(\alpha_s^2)$ corrections
to the processes $\eta_b(\chi_{bJ})\to J/\psi J/\psi$, which can provide useful guidance for experimental measurement. 

This paper is organized as follows. In Sec.~\ref{sec-general-formula} , we present the general formulas for the helicity amplitudes and (un)polarized decay 
widths of $\eta_b(\chi_{bJ})\to J/\psi J/\psi$. In Sec.~\ref{sec-nrqcd}, we factorize the helicity amplitudes by employing the NRQCD factorization.
In Sec.~\ref{sec-calculation}, we describe the technicalities encountered in the calculation 
and present the results of the various SDCs up to NNLO in $\alpha_s$. Sec.~\ref{sec-phen} is devoted to the phenomenological analysis and discussion. 
A brief summary is given in Sec.~\ref{sec-summary}. 
In Appendix~\ref{appendix-helicity-projectors}, we present the explicit expressions of the eight helicity projectors.
In Appendix~\ref{appendix-branching-ratios-mc}, the decay widths as well as the branching fractions at various level of accuracy are tabulated for different ratios of the charm quark mass 
and the bottom quark mass.

\section{(Un)polarized decay widths\label{sec-general-formula}}

It is convenient to apply the helicity amplitude formalism~\cite{Haber:1994pe,Jacob:1959at} to analyze the exclusive decay $H\to J/\psi J/\psi$, where $H$
can be $\eta_b$ or $\chi_{bJ}$.
We assign the magnetic number $S_z$ of the decaying particle directs along the $z$-axis and $\theta$ denotes the polar angle between $z$-axis and the direction of the outgoing $J/\psi$. Let $\lambda_1$ and $\lambda_2$ represent the helicities of the two outgoing $J/\psi$. 
The differential rate of $H$ decay into $J/\psi(\lambda_1)+J/\psi(\lambda_2)$ becomes 
%-----------------------------------
\beq\label{eq-gen-rate-helicity}
%-----------------------------------
\frac{d \Gamma\left[H\left(S_{z}\right) \rightarrow J / \psi(\lambda_1)J/\psi(\lambda_2)\right]}{d \cos \theta}=\frac{|\mathbf{P}|}{16 \pi m_{H}^{2}}\left|d_{S_{z}, \lambda_1-\lambda_2}^{J}(\theta)\right|^{2}\left|A_{\lambda_1, \lambda_2}^{H}\right|^{2},
%-----------------------------------
\eeq
%-----------------------------------
where $m_H$ represents the mass of $H$, and $\mathbf{P}$ denotes the spatial components of the $J/\psi$ momentum. 
The magnitude of  $\mathbf{P}$ is readily determined via
%-----------------------------------
\beq
%-----------------------------------
|{\bf P}|=\frac{\lambda^{1/2}(m_H^2,m_{J/\psi}^2,m_{J/\psi}^2)}{2m_H}=\sqrt{\frac{m_H^2}{4}-m_{J/\psi}^2},
%-----------------------------------
\eeq
%-----------------------------------
where the K\"allen function is defined via $\lambda(x,y,z)=x^2+y^2+z^2-2xy-2xz-2yz$. 
Note that angular momentum conservation constrains $|\lambda_{1}-\lambda_{2}|\leq J$, here $J$ is the spin of $H$.
The angular distribution is fully dictated by the quantum numbers $\lambda_1$ and $\lambda_2$  through the Wigner 
function $d_{S_{z}, \lambda_1-\lambda_2}^{J}(\theta)$, and 
$A_{\lambda_1, \lambda_2}^{H}$ is the intended helicity amplitude that encapsulates all nontrivial strong interaction dynamics, 
which depends upon $\lambda_1$ and $\lambda_2$.

Integrating (\ref{eq-gen-rate-helicity}) over $\cos\theta$ (one should cover only the hemisphere of the solid angle since two $J/\psi$ are indistinguishable
bosons),  and averaging over all possible $H$ polarizations, one obtains
%-----------------------------------
\begin{equation}\label{eq-gen-rate-helicity-int}
\begin{aligned}
\Gamma\left[H \rightarrow J/\psi(\lambda_1)J/\psi(\lambda_{2})\right]&=\frac{|\mathbf{P}|}{16 \pi m_{H}^{2}}\left|A_{\lambda_1, \lambda_2}^{H}\right|^{2} \int_{0}^{1}  \frac{1}{2J+1} \sum_{S_{z}}\left|d_{S_{z},\lambda_1-\lambda_2}^{J}(\theta)\right|^{2}d \cos \theta \\
&=\frac{|\mathbf{P}|}{16\pi(2J+1) m_{H}^{2}}\left|A_{\lambda_1,\lambda_2}^{H}\right|^{2}.
\end{aligned}
\end{equation}
%----------------------------

The helicity amplitudes are not independent.
Due to the parity invariance~\cite{Haber:1994pe}, there are following relations
%-----------------------------------
\begin{equation}\label{eq-helicity-parity-invariance}
\begin{aligned}
A^{\eta_{b}}_{\lambda_1,\lambda_2}=-A^{\eta_{b}}_{-\lambda_1,-\lambda_2},\quad 
A^{\chi_{bJ}}_{\lambda_1,\lambda_2}=(-1)^J A^{\chi_{bJ}}_{-\lambda_1,-\lambda_2}.
\end{aligned}
\end{equation}
%----------------------------
Moreover, we have 
%-----------------------------------
\begin{equation}\label{eq-helicity-identical}
\begin{aligned}
A^{\eta_{b}}_{\lambda_1,\lambda_2}=-A^{\eta_{b}}_{\lambda_2,\lambda_1},  \quad A^{\chi_{bJ}}_{\lambda_1,\lambda_2}=(-1)^J A^{\chi_{bJ}}_{\lambda_2,\lambda_1}
\end{aligned}
\end{equation}
%----------------------------
for the two identical $J/\psi$ in the final state~\cite{Haber:1994pe}.
Thus, there are one independent helicity amplitude for $\eta_{b}$ and $\chi_{b1}$ decay,  two for $\chi_{b0}$, and four for $\chi_{b2}$.

It is straightforward to obtain the unpolarized 
decay rates for $H\to J/\psi J/\psi$ by summing over all the allowed helicity channels
%-----------------------------------
\begin{subequations}\label{eq-gen-rate-helicity-explicit}
	%---------------------------------------------
	\begin{eqnarray}
	%---------------------------------------------
	\Gamma(\eta_b\to J/\psi J/\psi)&=&
	\frac{|{\bf P}|}{16\pi m_{\eta_b}^{2}}\bigg(2| A_{1,1}^{\eta_{b}}|^2\bigg),\\
	%---------------------------------------------
	\Gamma(\chi_{b0}\to J/\psi J/\psi)&=&
	\frac{|{\bf P}|}{16\pi m_{\chi_{b0}}^{2}}\bigg(2|A_{1,1}^{\chi_{b0}}|^2+|A_{0,0}^{\chi_{b0}}|^2\bigg),\\
	%---------------------------------------------
	\Gamma(\chi_{b1}\to J/\psi J/\psi)&=&
	\frac{|{\bf P}|}{48\pi m_{\chi_{b1}}^{2}}\bigg(4| A_{1,0}^{\chi_{b1}}|^2\bigg),\\
	%---------------------------------------------
	\Gamma(\chi_{b2}\to J/\psi J/\psi)&=&
	\frac{|{\bf P}|}{80\pi m_{\chi_{b2}}^{2}}\bigg(2| A_{1,-1}^{\chi_{b2}}|^2
	+2|A_{1,1}^{\chi_{b2}}|^2
	+4|A_{1,0}^{\chi_{b2}}|^2+|A_{0,0}^{\chi_{b2}}|^2\bigg).
	%---------------------------------------------
	\end{eqnarray}
	%---------------------------------------------
\end{subequations}
%-----------------------------------

Finally, it is enlightening to analyze the asymptotic behavior of the helicity amplitudes.  
In the limit of $m_b\gg m_c$,  $A_{\lambda_{1},\lambda_{2}}^{H}$ satisfies
\begin{equation}\label{eq-helicity-scaling-rule-2}
A_{\lambda_1, \lambda_2}^H \propto r^{2+\left|\lambda_1+\lambda_2\right|}
\end{equation}
where $r=\frac{m_c}{m_b}$. 
For each $J/\psi$ production,  it contributes one power of $r$ in (\ref{eq-helicity-scaling-rule-2})
originating from the large momentum transfer which is 
required for the charm-anticharm pair to form the $J/\psi$ with small relative momentum.  
The other powers of $r$ arise from the helicity selection rule in
perturbative QCD~\cite{Chernyak:1980dj,Brodsky:1981kj}.

%-----------------------------------
\section{NRQCD factorization for the helicity amplitude \label{sec-nrqcd}}
%-----------------------------------

By employing the NRQCD factorization~\cite{Bodwin:1994jh}, we can express the helicity amplitude into
%-----------------------------------
\bqa\label{eq-nrqcd}
%-----------------------------------
A^{H}_{\lambda_1,\lambda_2}=\sqrt{2m_H}2m_{J/\psi}f_{\lambda_1,\lambda_2}^H
\frac{\sqrt{\langle \mathcal{O}\rangle_H}}{m_b^{n}} \frac{\langle\mathcal{O}\rangle_{J/\psi}}{m_c^3},
%-----------------------------------
\eqa
%-----------------------------------
where $n=2$ for $\eta_b$ and $n=3$ for $\chi_{bJ}$,  and
$f_{\lambda_1,\lambda_2}^H$ denotes the dimensionless short-distance coefficient (SDC). The nonperturbative long-distance matrix element (LDME) is defined via
$\langle \mathcal{O} \rangle_H\equiv |\langle 0|\mathcal{O}_{H} |H\rangle|^2$ with
%--------------------------------------------
\begin{subequations}
\begin{eqnarray}
{\mathcal O}_{\eta_b}&=& \chi^{\dagger}\psi,\\
{\mathcal O}_{\chi_{b0}}&=&\frac{1}{\sqrt{3}}\chi^{\dagger}\left(-\frac{i}{2}\overleftrightarrow{{\bf
D}}\cdot {\bm\sigma}\right)\psi,\\
{\mathcal O}_{\chi_{b1}}&=&\frac{1}{\sqrt{2}}\chi^{\dagger}\left(-\frac{i}{2}\overleftrightarrow{{\bm
D}}\times{\bm \sigma}\right)\cdot {\bm \epsilon}_{\chi_{b1}}\psi,\\
{\mathcal O}_{\chi_{b2}}&=&\chi^{\dagger}\left(-{i\over 2}\overleftrightarrow{D}^{(i}\sigma^{j)}\right)\epsilon_{\chi_{b2}}^{ij}\psi,\\
{\mathcal O}_{J/\psi}&=& \chi^{\dagger}{\bm \sigma}\cdot {\bm \epsilon}_{J/\psi}\psi.
\end{eqnarray}
\end{subequations}
%--------------------------------------------
where $\psi$  and $\chi^{\dagger}$ are the Pauli spinor fields annihilating a heavy quark and antiquark respectively, and $\epsilon_{H}$ and $\epsilon_{J/\psi}$
represent the polarization tensor/vector of $H$ and $J/\psi$ respectively.

The prefactor $\sqrt{2m_H}2m_{J/\psi}$ in (\ref{eq-nrqcd}) originates from the fact that we adopt relativistic normalization for the quarkonium in the helicity amplitude, however, we adopt nonrelativistic normalization in the LDMEs. Since, in this work, we are only concerned with the lowest order in velocity 
expansion, we can set $m_H=2m_b$ and $m_{J/\psi}=2 m_c$ in (\ref{eq-nrqcd}).

It worth noting that the helicity selection rule for SDCs
%-----------------------------------
\bqa\label{eq-helicity-scaling-rule-1}
%-----------------------------------
f_{\lambda_1,\lambda_2}^{H}\propto r^{1+|\lambda_1+\lambda_2|}
%-----------------------------------
\eqa
%-----------------------------------
can be directly deduced from (\ref{eq-helicity-scaling-rule-2}) by noticing that $\langle\mathcal{O}\rangle_{J/\psi}\propto m_c^{3}$.

It is convenient to expand the helicity SDCs in powers of the strong coupling constant
\bqa\label{eq-sdcs-expand-qcd}
%-----------------------------
f_{\lambda_1, \lambda_2}^{H}&=&\alpha_s^2\Big[f_{\lambda_1, \lambda_2}^{{H},(0)}+\frac{\alpha_s}{\pi} \left(\frac{\beta_0}{2} \ln \frac{\mu_R^2}{m_b^2} f_{\lambda_1, \lambda_2}^{{H},(0)}+f_{\lambda_1, \lambda_2}^{{H},(1)}\right)+\frac{\alpha_s^2}{\pi^2} \Big(\frac{3\beta_0^2}{16} \ln^2 \frac{\mu_R^2}{m_b^2} f_{\lambda_1, \lambda_2}^{{H},(0)}+\big(\frac{\beta_1}{8}f_{\lambda_1, \lambda_2}^{{H},(0)}\nn\\
&+&\frac{3\beta_0}{4}f_{\lambda_1, \lambda_2}^{{H},(1)}\big) \ln \frac{\mu_R^2}{m_b^2}+\big(2\gamma_{J/\psi}+\gamma_H\big) \ln \frac{\mu_\Lambda^2}{m_c^2}f_{\lambda_1, \lambda_2}^{{H},(0)}+f_{\lambda_1, \lambda_2}^{{H},(2)}\Big)\Big]+\mathcal{O}\left(\alpha_s^5\right),
\eqa
%-----------------------------
where $\beta_0=\frac{11}{3}C_A-\frac{4}{3}T_F n_f$ and $\beta_1=\frac{34}{3}C_A^2-(\frac{20}{3}C_A+4C_F)T_F n_f$ are the one-loop  and two-loop coefficients of the QCD $\beta$ function respectively, and $n_f=n_L+n_H$ signifies the number of the active flavors, with the number of light quarks $n_L$=3, and the number of heavy quark $n_H=1$.
$\mu_R$ and $\mu_\Lambda$ refer to the renormalization scale and NRQCD factorization scale, respectively.
The $\ln \mu_R^2$ terms in (\ref{eq-sdcs-expand-qcd}) guarantee the renormalization group invariance of the SDCs.
The occurrence of $\ln \mu_\Lambda^2$ is required by the NRQCD factorization.
According to the factorization, the $\mu_\Lambda$ dependence in the SDCs should be thoroughly eliminated by that in the LDMEs. 
 $\gamma_{J/\psi}$ and $\gamma_{H}$ represent the anomalous dimensions associated with the NRQCD bilinear currents carrying the quantum numbers $^3S_1$,  $^1S_0$ or $^3P_J$, which have already been known from various sources~\cite{Czarnecki:1997vz,Beneke:1997jm,Czarnecki:2001zc,Hoang:2006ty,Sang:2015uxg,Sang:2020fql}
%-----------------------------
\begin{subequations}\label{eq-ano-dim}
	\bqa
	%-----------------------------
	\gamma_{J/\psi}&=&-\pi^2\left(\frac{C_A C_F}{4}+\frac{C_F^2}{6}\right),\\
	%-------------------------
	\gamma_{\eta_b}&=&-\pi^2\left(\frac{C_A C_F}{4}+\frac{C_F^2}{2}\right),\\
	%-------------------------
	\gamma_{\chi_{b0}}&=&-\pi^2\left(\frac{C_A C_F}{12}+\frac{C_F^2}{3}\right),\\
	%-------------------------
	\gamma_{\chi_{b1}}&=&-\pi^2\left(\frac{C_A C_F}{12}+\frac{5C_F^2}{24}\right),\\
	%-----------------------------
	\gamma_{\chi_{b2}}&=&-\pi^2\left(\frac{C_A C_F}{12}+\frac{13C_F^2}{120}\right).
	\eqa
\end{subequations}
%-----------------------------

The SDCs can be determined by the perturbative matching procedure, i.e., by replacing the physical $J/\psi/H$ with the fictitious quarkonia composed of the free $c\bar{c}/b\bar{b}$ pair with the same quantum numbers as $J/\psi/H$, computing both sides in (\ref{eq-nrqcd}) in perturbative QCD and NRQCD, then solving for the SDCs order by order in perturbation theory.

%-----------------------------
\section{SDCS up to $\mathcal{O}(\alpha_s)$ \label{sec-calculation}}
%-----------------------------

%----------------------------
\begin{figure}[htbp]
	\centering
	\includegraphics[width=1\textwidth]{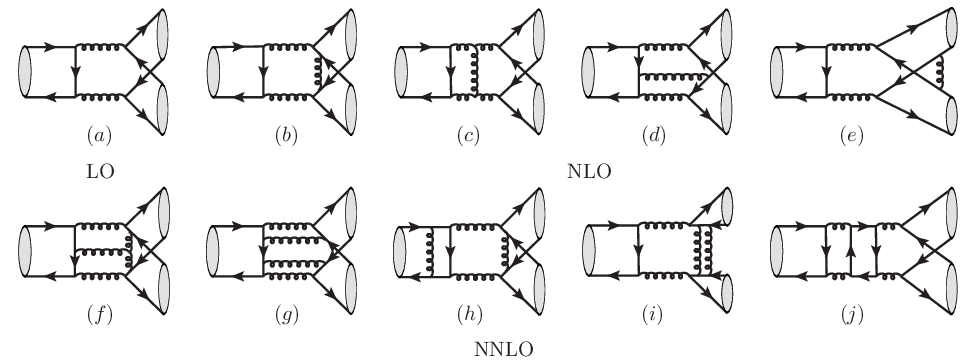}
	\caption{Some representative Feynman diagrams for the process $H\to J/\psi J/\psi$ up to two loop.
		\label{fig-feynman-diagram}}
\end{figure}
%----------------------------

We employ the {\tt FeynArts}~\cite{Hahn:2000kx} to generate the quark-level Feynman diagrams and Feynman amplitudes for $b\bar{b}\to c\bar{c}+c\bar{c}$.
We adopt the well-known covariant color/spin/orbital projector technique with the aid of the packages {\tt FeynCalc}~\cite{Shtabovenko:2016sxi} and {\tt FormLink}~\cite{Feng:2012tk} to expedite the matching calculation.
To further extract the helicity amplitudes, we find it convenient to apply various covariant helicity projectors.
The expressions of these helicity projectors are explicitly presented in Appendix~\ref{appendix-helicity-projectors}.

It is straightforward to compute the leading ordr (LO) SDCs
%-----------------
\begin{subequations}\label{lo-sdc}
	\bqa
	f^{\chi_{b0},(0)}_{1,1}&=&\frac{64\sqrt{3}\pi^2 r^3}{81},\qquad
	f^{\chi_{b0},(0)}_{0,0}=-\frac{32\sqrt{3}\pi^2r(1-2r^2)}{81},\\
	f^{\chi_{b2},(0)}_{1,1}&=&-\frac{64\sqrt{6}\pi^2r^3}{81},\qquad
	f^{\chi_{b2},(0)}_{1,0}=\frac{32\sqrt{2}\pi^2r^2}{27},\\
	f^{\chi_{b2},(0)}_{1,-1}&=&-\frac{32\pi^2r}{27},\qquad
	f^{\chi_{b2},(0)}_{0,0}=-\frac{16\sqrt{6}\pi^2r(1+4r^2)}{81},
	\eqa
\end{subequations}
%-----------------
which are consistent with those in Refs~\cite{Zhang:2011ng,Sun:2010qx,Gong:2008ue}.  
Accidentally, we find the SDCs for $\eta_b$ and $\chi_{b1}$ vanish at LO. 
This can be explained by the following fact.
At LO, the amplitude of  the subprocess $\eta_{b}(\chi_{b1})\to g^*g^*$ is proportional to a Levi-Civita tensor, and
by coincidence,  the momenta of the two virtual gluons are equal, which cause the Levi-Civita tensor to vanish.   
In addition, it is straightforward to check that the $f^{H,(0)}$ in (\ref{lo-sdc}) satisfy the helicity scaling rule (\ref{eq-helicity-scaling-rule-1}) in the limit of  $r\to 0$.

Once beyond the LO, we adopt the standard shortcut to directly extract the SDCs, i.e., 
to expand the QCD amplitudes in powers of quark relative momentum prior to conducting loop integrals, 
which amounts to directly extracting the contribution from
the hard region in the context of strategy of region~\cite{Beneke:1997zp}.  
We utilize the packages {\tt Apart}~\cite{Feng:2012iq} and {\tt FIRE}~\cite{Smirnov:2014hma}
to reduce the loop integrals into linear combinations of  a group of master integrals (MIs).
Finally, we end up with 20 one-loop MIs, which are analytically computed with the aid of the {\tt Package-X}~\cite{Patel:2015tea},
and 1439 two-loop MIs, which are numerically evaluated with the method of Auxiliary Mass Flow~\cite{Liu:2017jxz,Liu:2022mfb,Liu:2021wks,Liu:2022chg}. Moreover, we employ the newly released package {\tt CalcLoop}~\cite{Ma:CalcLoop},  developed by Ma {\it et al.}, to check some of our computation. 

Performing the field-strength and mass renormalization, with two-loop expressions of $Z_2$ and $Z_m$ taken from~\cite{Broadhurst:1991fy}, and
renormalizing the strong coupling constant under the $\overline{\rm MS}$ scheme to two-loop order, we eliminate the UV divergences
in the two-loop SDCs.
Nevertheless, the renormalized two-loop corrections to the SDCs still contain uncancelled
single IR poles.
This pattern is exactly what is required by NRQCD
factorization as reflected in (\ref{eq-sdcs-expand-qcd}). 
These IR poles can be factored into the NRQCD LDMEs under the $\overline{\rm MS}$ prescription, which then become scale-dependent quantities.
As mentioned previously,  the $\ln \mu_\Lambda$ terms in (\ref{eq-sdcs-expand-qcd}) exactly cancel the $\mu_\Lambda$ dependence of the LDMEs, so that the 
predicted decay width is independent of $\mu_\Lambda$.  The validity of NRQCD factorization in this
process turns out to be highly nontrivial.

Since the analytic expressions of the $f^{H,(1)}_{\lambda_1,\lambda_2}$ are too complicated,  here we merely present their asymptotic expansion in the limit of $r\to 0$
%-----------------------------
	\bqa
	f^{H,(1)}_{\lambda_1,\lambda_2}\bigg|_{\rm asy}=\frac{64 \pi^2}{81}r^{1+|\lambda_1+\lambda_2|}\mathcal{C}^{H}_{\lambda_1,\lambda_2},
	\eqa
%----------------------------
where we have deliberately pulled out the $r$-dependent factor in front to make the helicity scaling rule manifest, 
so that the $\mathcal{C}^{H}_{\lambda_1,\lambda_2}$ scale as $r^0$ and read
%------------------------------
\begin{subequations}\label{nlo-sdc}
	\bqa
	%-----------------------------
	\mathcal{C}^{\eta_b}_{1,1}&=&\frac{19 \ln ^2r}{8}+\left(\frac{5}{2}+\frac{19 i \pi }{8}-\frac{\ln2}{4}\right) \ln r+\frac{5 \ln ^22}{16}
	+\left(\frac{1}{2}+\frac{i \pi }{8}\right) \ln 2+\frac{29 \pi ^2}{96}-\frac{3 \sqrt{3} \pi \
	}{8}\nn\\
	&+&i\pi+\frac{3}{4},\\
	%-----------------------------------------
	\mathcal{C}^{\chi_{b0}}_{1,1}&=&-\frac{59 \ln ^2r}{8 \sqrt{3}}+\sqrt{3}\left(\frac{25 \ln 2}{12}-\frac{22}{3}-\frac{59 i \pi }{24}\right)\ln r+\frac{25\sqrt{3}}{16} \ln ^22-\sqrt{3} \left(\frac{11}{2}+\frac{33 i \pi }{8}\right) \ln 2 \nn\\
	&+&\sqrt{3} \left(\frac{983}{144}+\frac{65 i\pi }{24}-\frac{19 \pi ^2}{32}\right)+\frac{\pi }{3}-\frac{\sqrt 3 n_f}{9}\left(3i\pi+5\right)
	,\\
	%------------------------------------------------
	\mathcal{C}^{\chi_{b0}}_{0,0}&=&\frac{(15-10 \ln 2)\sqrt{3}}{6} \ln r-\sqrt{3} \
	\left(\frac{1}{2}+\frac{13 i \pi }{24}\right) \ln 2+\sqrt{3} \left(\frac{17 \pi^2}{96}-\frac{511}{144}-\frac{25 i \pi }{24}\right)
	\nn\\
	&+&\frac{\sqrt{3}}{4} \ln ^22+\frac{37 \pi }{48}+\frac{\sqrt 3 n_f}{18}\left(3i\pi+5\right),\\
	%----------------------------------------------------
	\mathcal{C}^{\chi_{b1}}_{1,0}&=&\frac{\ln ^2r}{8 \sqrt{2}}+\sqrt{2}\left(\frac{5 \ln 2}{8}-\frac{7}{64}+\frac{i \pi }{16}\right) \ln r-\frac{97 \ln^2 2}{8 \sqrt{2}}+\sqrt{2} \left(\frac{159}{16}+\frac{51 i \pi }{4}\right) \ln 2\nn\\
	&+&\sqrt{2} \left(\frac{13 \pi ^2}{96}-\frac{583}{256}-\frac{127 i \pi }{16}-\frac{21 \sqrt{3} \pi}{32}\right),\\
	%----------------------------------------------
	\mathcal{C}^{\chi_{b2}}_{1,1}&=&\frac{13 \ln^2r}{2 \sqrt{6}}-\sqrt{6}\left(\frac{37 \ln 2}{12}-\frac{29}{6}-\frac{13 i \pi }{12}\right) \ln r+\frac{411\sqrt 6}{16} \ln^2 2-\sqrt{6} \left(\frac{1123}{30}+52 i \pi\right) \ln 2\nn\\
	&+&\sqrt{6} \left(\frac{3851}{720}+\frac{731 i \pi }{20}+\frac{29 \pi^2}{96}\right)+\frac{1261 \pi }{120 \sqrt{2}}+\frac{\sqrt{6}n_f}{9}\left(3i\pi+5\right)
	,\\ 
	%-------------------------------------------------
	\mathcal{C}^{\chi_{b2}}_{1,0}&=&-\frac{17 \ln ^2r}{8 \sqrt{2}}+\sqrt{2}\left(\frac{21 \ln 2}{8}-\frac{335}{64}-\frac{17 i \pi }{16}\right) \ln r+\frac{417 \ln^22}{8\sqrt{2}}-\sqrt{2} \left(\frac{8879}{240}+\frac{857 i \pi }{16}\right) \ln 2\nn\\
	&+&\sqrt{2} \left(\frac{76087}{3840}+\frac{17779 i \ \pi }{480}-\frac{67 \pi ^2}{64}\right)-\frac{181\sqrt{6} \pi }{480}-\frac{\sqrt{2}n_f}{6} \left(3i\pi+5\right)
	,\\ 
	%-----------------------------------------------------
	\mathcal{C}^{\chi_{b2}}_{1,-1}&=&\left(6-8 \ln 2\right) \ln r+10 \ln
	^22-\left(\frac{453}{20}+\frac{163 i \pi }{8}\right) \ln 2+\frac{73 \pi^2}{96}+\frac{5801 i \pi }{480}-\frac{1157}{240}\nn\\
	&+&\frac{1081 \sqrt{3}\pi }{480}+\frac{n_f}{6} \left(3i\pi+5\right),\\
	%-------------------------------------------
	\mathcal{C}^{\chi_{b2}}_{0,0}&=&\frac{\left(3-2\ln
		2\right)\sqrt 6}{6} \ln r+\frac{13\sqrt 6}{4} \ln
	^22-\sqrt{6} \left(\frac{259}{60}+\frac{161 i \pi }{24}\right) \ln 2\nn\\
	&-&\sqrt{6} \left(\frac{1067}{1440}-\frac{2141 i \pi}{480}+\frac{\pi ^2}{32}\right)+\frac{197\sqrt 2 \pi}{120}+\frac{\sqrt{6}n_f}{36}
	\left(3i\pi+5\right).
	\eqa
\end{subequations}
%-----------------------------

Finally we remain to identify the desired nonlogarithmic piece in the two-loop SDCs. 
It becomes much more challenging to deduce the analytical expressions for the encountered two-loop MIs. 
Instead, we resort to the high-precision numerical computation.
By taking the bottom quark and charm quark pole masses to the typical values $m_b = 4.70\, {\rm GeV}$ and $m_c=1.50\, {\rm GeV}$ respectively,
we tabulate the results of $f^{H,(2)}_{\lambda_1,\lambda_2}$
in Table~\ref{tab-sdcs-mc-1.5}. For completeness, the numerical values of $f^{H,(0)}_{\lambda_1,\lambda_2}$ and $f^{H,(1)}_{\lambda_1,\lambda_2}$ are also listed.

%-----------------------------------------------------------
	\begin{table}[!htbp]\normalsize
			\caption{Numerical values for various SDCs with $m_b=4.7$ GeV and $m_c=1.5$ GeV.}
			\label{tab-sdcs-mc-1.5}
			\centering
			\setlength{\tabcolsep}{20pt}
			\renewcommand{\arraystretch}{1.6}
			%---------------------------------
			%------------------------------------
			\begin{tabular}{|c|c|c|c|c|}
				\hline
				\multicolumn{5}{|l|}{$m_{c}=1.50\, {\rm GeV},m_{b}=4.70\, {\rm GeV}$}\\
				\hline
				H&($\lambda_1,\lambda_2$)&$f_{\lambda_1,\lambda_2}^{(0)}$&$f_{\lambda_1,\lambda_2}^{(1)}$&$f_{\lambda_1,\lambda_2}^{(2)}$\\
				\hline
				$\eta_{b}$
				&(1,1)&--&$0.551-1.410i$&$21.624-10.293i$ \\
				%----------------------------------------
				\hline
				\multirow{2}*{$\chi_{b0}$} 
				&(1,1)&$0.439$&$-0.266+2.726i$&$-86.938+29.858i$ \\
				%----------------------------------------
				\cline{2-5}
				&(0,0)&$-1.716$&$-2.830-6.283i$&$184.697-75.742i$ \\
				%----------------------------------------
				\hline
				$\chi_{b1}$
				&(1,0)&--&$-0.281+1.188i$&$8.812+7.876i$ \\
				%----------------------------------------
				\hline
				\multirow{4}*{$\chi_{b2}$} 
				&(1,1)&$-0.621$&$3.801+0.233i$&$84.482-2.120i$ \\
				%----------------------------------------
				\cline{2-5}
				&(1,0)&$1.685$&$-9.666-1.178i$&$-221.561-1.457i$ \\
				%----------------------------------------
				\cline{2-5}
				&(1,-1)&$-3.733$&$21.499+2.214i$&$465.006+10.900i$ \\
				\cline{2-5}
				&(0,0)&$-2.145$&$11.326+2.567i$&$278.731+3.346i$ \\
				\hline
				\end{tabular}
				\end{table}
%--------------------------------------------------------------------
\section{Phenomenology and discussion\label{sec-phen}}

Prior to making phenomenological predictions, we specify our choice of the various input parameters. 
In order to reduce the theoretical uncertainty, we use the physical quarkonium masses in computing the phase space in (\ref{eq-gen-rate-helicity-explicit}). 
But beyond that, we choose $m_H=2m_b$ and $m_{J/\psi}=2m_c$ so as to maintain gauge invariance.
The physical quarkonium mass are taken from the particle data group (PDG)\cite{ParticleDataGroup:2020ssz}:
$m_{\eta_b}=9.3987$ GeV, $m_{\chi_{b0}}=9.85944$ GeV, $m_{\chi_{b1}}=9.89278$ GeV, $m_{\chi_{b2}}=9.91221$ GeV and $m_{J/\psi}=3.0969$ GeV. 
We choose the benchmark values of the heavy quark pole masses $m_b=4.7$ GeV and $m_c=1.5$ GeV. In addition, we will investigate the dependence of the theoretical results on the mass ratio $r$.  

We approximate the NRQCD LDMEs at $\mu_\Lambda=1$ GeV  by the Schr\"odinger radial wave function at the origin and the
first derivative of the Schr\"odinger radial wave function at the origin for the S-wave and P-wave quarkonium, respectively,
%-----------------------------
\begin{subequations}\label{eq-hqs}
	%-----------------------------
	\bqa
	\left\langle\mathcal{O}\right\rangle_{J/\psi} & \approx \frac{N_c}{2 \pi}\left|R_{1 S}^{c \bar{c}}(0)\right|^2=\frac{N_c}{2 \pi} \times 0.810 \mathrm{GeV}^3,\\
	%-------------------------
	\left\langle\mathcal{O}\right\rangle_{\eta_b} & \approx \frac{N_c}{2 \pi}\left|R_{1 S}^{b \bar{b}}(0)\right|^2=\frac{N_c}{2 \pi} \times 6.477 \mathrm{GeV}^3 ,\\
	%-----------------------------
	\left\langle\mathcal{O}\right\rangle_{\chi_{bJ}} & \approx \frac{3 N_c}{2 \pi}\left|R_{1 P}^{\prime b \bar{b}}(0)\right|^2=\frac{3 N_c}{2 \pi} \times 1.417 \mathrm{GeV}^5,
	\eqa
\end{subequations}
%-----------------------------
where the radial wave functions at the origin are evaluated from  Buchm\"uller-Tye (BT) potential model~\cite{Eichten:1995ch}.
Note that we have made the approximation $\langle\mathcal{O}\rangle_{\chi_{b0}}\approx \langle\mathcal{O}\rangle_{\chi_{b1}}\approx \langle\mathcal{O}\rangle_{\chi_{b2}}$ by invoking the heavy quark spin symmetry.

We fix $\mu_\Lambda=1$ GeV. The central value of $\mu_R$ is chosen $\mu_R=m_b$ and we vary $\mu_R$ from $2m_c$ to $2m_b$ to estimate the theoretical
uncertainties.  The QCD running coupling constant is evaluated with the aid of the package {\tt RunDec}~\cite{Chetyrkin:2000yt} at two loop.

	%-----------------------------------------------------------
\begin{table}[!htbp]\normalsize
	\caption{Total decay widths of $\chi_{bJ}$.}
	\label{total-widths}
	\centering
	\setlength{\tabcolsep}{23pt}
	\renewcommand{\arraystretch}{2}
%-----------------------------------------------------------
	\begin{tabular}{cccccc}
		\hline
		\hline
		$H$ & ${\rm \Gamma}[\chi_{bJ}\to \gamma\Upsilon ]$(keV)\cite{QuarkoniumWorkingGroup:2004kpm}&${\rm Br}[\chi_{bJ}\to \gamma\Upsilon]$\cite{ParticleDataGroup:2020ssz}&$\Gamma_{tot}$(MeV) \\
		%----------------------------------------
		\hline
		$\chi_{b0}$&22.2&(1.94$\pm$0.27)\%&$1.144^{+0.185}_{-0.140}$ \\
		$\chi_{b1}$&27.8&(35.2$\pm$2.0)\%&$0.079^{+0.005}_{-0.004}$ \\
		$\chi_{b2}$&31.6&(18.0$\pm$1.0)\%&$0.176^{+0.010}_{-0.009}$ \\
		%----------------------------------------
		\hline
		\hline
		%----------------------------------------
	\end{tabular}
\end{table}
%-------------------------------------------------------

To further predict the branching fractions of various decay channels, we need to specify the total decay widths of $\eta_b$ and $\chi_{bJ}$. 
The decay width of $\eta_b$ can be directly taken from the PDG~\cite{ParticleDataGroup:2020ssz}: $\Gamma_{\eta_b}=10^{+5}_{-4}$ MeV.
So far, the decay widths of $\chi_{bJ}$ are absent from the PDG.
Nevertheless,
we can determine the decay widths of $\chi_{bJ}$ through
 %-----------------------------
	\bqa
   \Gamma_{tot}(\chi_{bJ})=\frac{{\rm \Gamma}[\chi_{bJ}\to \gamma\Upsilon ]}{{\rm Br}[\chi_{bJ}\to \gamma\Upsilon]},
	\eqa
%-----------------------------
where the branching fractions of the E1 transition are measured~~\cite{ParticleDataGroup:2020ssz}, and
 the decay widths of the E1 transition have been given in Ref.~\cite{QuarkoniumWorkingGroup:2004kpm}.
We enumerate all the results in Table~\ref{total-widths}, where the uncertainty in $\Gamma_{tot}$ originates from the ${\rm Br}[\chi_{bJ}\to \gamma\Upsilon]$.

Now, we collect all the ingredients to perform phenomenological analysis. 
In Table~\ref{tab-widths}, we tabulate 
the theoretical predictions on the (un)polarized decay widths and the branching fractions at various levels of
accuracy in $\alpha_s$. To facilitate comparison,  the upper limits of various channels from the {\tt Belle} measurements~\cite{Shen:2012ei} 
are listed in the last column.   The uncertainty affiliated with the decay width is caused by sliding the renormalization scale $\mu_R$, and the two uncertainties in the branching fraction are from the renormalization scale and total decay width. 
We should emphasize that there are other sources of uncertainties for the theoretical predictions, e.g, the values of the Schr\"odinger wave functions and
the uncalculated relativistic corrections, which may potentially bring about extra uncertainties.
In addition, we do not include the contributions from the Feynman diagrams where 
the double $J/\psi$ are produced through two virtual photon independent fragmentation, which actually are much less than the nonfragmentation contributions~\cite{Jia:2006rx,Gong:2008ue,Zhang:2011ng,Chen:2012ih} and therefore can be safely neglected. 
It worth noting that the situation is quite different from the process $e^+e^-\to J/\psi J/\psi$, where 
the dominant production mechanism is via two photon independent
fragmentation into $J/\psi$.

%------------------------------------------------------------------------
\begin{table}[!htbp]\small
	\caption{Theoretical predictions on various (un)polarized decay widths (in units of eV) and branching fractions ($\times 10^{-5}$).}
	\label{tab-widths}
	\centering
	\setlength{\tabcolsep}{1pt}
	\renewcommand{\arraystretch}{1.8}
	%---------------------------------
	%------------------------------------
	\begin{tabular}{|c|c|c|c|c|c|c|c|c|}
		\hline
		H&Order&$\Gamma_{0,0}$&$\Gamma_{1,0}$&$\Gamma_{1,1}$&$\Gamma_{1,-1}$&$\Gamma_{\rm Unpol}$&$\rm Br_{th}$&$\rm Br_{exp}$\cite{Shen:2012ei}\\
		\hline
		\multirow{3}*{$\eta_{b}$}
		&LO&--&--&--&--&--&--&\multirow{3}*{--}\\
		\cline{2-8}
		&NLO&--&--&$1.080^{+1.663}_{-0.714}$&--&$2.160^{+3.325}_{-1.428}$&$0.022^{+0.033+0.014}_{-0.014-0.007}$&\\
		%----------------------------------------
		\cline{2-8}
		&NNLO&--&--&$4.084^{+3.987}_{-2.232}$&--&$8.168^{+7.973}_{-4.463}$&$0.082^{+0.080+0.054}_{-0.045-0.027}$& \\
		%----------------------------------------
		\hline
		\multirow{3}*{$\chi_{b0}$}
		&LO&$8.542^{+7.358}_{-4.393}$&--&$0.559^{+0.482}_{-0.288}$&--&$9.660^{+8.321}_{-4.968}$&$0.844^{+0.727+0.117}_{-0.434-0.118}$&\multirow{3}*{$\textless$7.1}\\
		%----------------------------------------
		\cline{2-8}
		&NLO&$11.140^{+1.233}_{-2.500}$&--&$0.616^{+0.084}_{-0.124}$&--&$12.372^{+1.400}_{-2.748}$&$1.081^{+0.122+0.150}_{-0.240-0.151}$& \\
		%----------------------------------------
		\cline{2-8}
		&NNLO&$6.449^{+1.710}_{-1.955}$&--&$0.329^{+0.371}_{-0.012}$&--&$7.107^{+1.741}_{-1.212}$&$0.621^{+0.152+0.086}_{-0.106-0.086}$& \\
		%----------------------------------------
		\hline
		\multirow{3}*{$\chi_{b1}$}
		&LO&--&--&--&--&--&--&\multirow{3}*{$\textless$2.7}\\
		\cline{2-8}
		&NLO&--&$0.007^{+0.011}_{-0.005}$&--&--&$0.027^{+0.042}_{-0.018}$&$0.035^{+0.053+0.002}_{-0.023-0.002}$&\\
		%----------------------------------------
		\cline{2-8}
		&NNLO&--&$0.014^{+0.009}_{-0.006}$&--&--&$0.057^{+0.035}_{-0.026}$&$0.072^{+0.044+0.004}_{-0.033-0.004}$& \\
		%----------------------------------------
		\hline
		\multirow{3}*{$\chi_{b2}$}
		&LO&$2.663^{+2.294}_{-1.370}$&$1.643^{+1.416}_{-0.845}$&$0.223^{+0.192}_{-0.115}$&$8.067^{+6.949}_{-4.149}$&$25.818^{+22.239}_{-13.279}$&$14.669^{+12.636+0.884}_{-7.545-0.789}$&\multirow{3}*{$\textless$4.5}\\
		%----------------------------------------
		\cline{2-8}
		&NLO&$1.094^{+0.281}_{-0.679}$&$0.604^{+0.199}_{-0.423}$&$0.075^{+0.030}_{-0.057}$&$2.943^{+0.987}_{-2.084}$&$9.545^{+3.111}_{-6.655}$&$5.424^{+1.768+0.327}_{-3.781-0.292}$& \\
		%----------------------------------------
		\cline{2-8}
		&NNLO&$0.071^{+0.476}_{-0.048}$&$0.020^{+0.267}_{-0.015}$&$0.001^{+0.032}_{-0.001}$&$0.157^{+1.351}_{-0.130}$&$0.467^{+4.311}_{-0.360}$&$0.265^{+2.450+0.016}_{-0.205-0.014}$& \\
		%----------------------------------------
		\hline
	\end{tabular}
\end{table}
%------------------------------------------------------------------------
Examining Table~\ref{tab-widths} closely, we find that the polarized decay widths roughly obey the hierarchy as indicated by the helicity scaling rule in (\ref{eq-helicity-scaling-rule-2}).
It is interesting to note that both the NLO and NNLO perturbative corrections to $\Gamma[\chi_{b2}\to J/\psi J/\psi]$
 are sizable and negative.  
Incorporating the perturbative corrections significantly reduces the LO prediction, which indicates that the perturbative convergence is rather poor. 
Because of this, the theoretical prediction bears large $\mu_R$ dependence for $\chi_{b2}$.
In addition, the decay widths of $\eta_b$ and $\chi_{b1}$ are much smaller than the other two channels, which is attributed to the vanishing LO amplitudes for 
$\eta_b$ and $\chi_{b1}$. 
Finally, we find the theoretical predictions on the branching fractions are consistent with the upper limits measured by the {\tt Belle} Collaboration~\cite{Shen:2012ei}.

\begin{figure}[htbp]
	\centering
	\includegraphics*[scale=0.95]{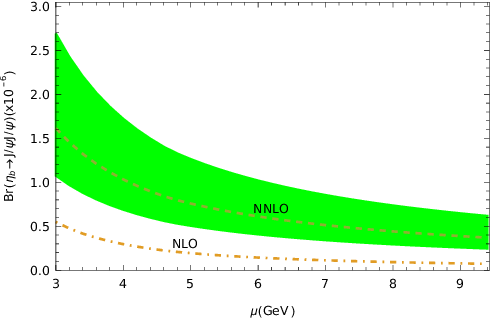}
	\includegraphics*[scale=0.95]{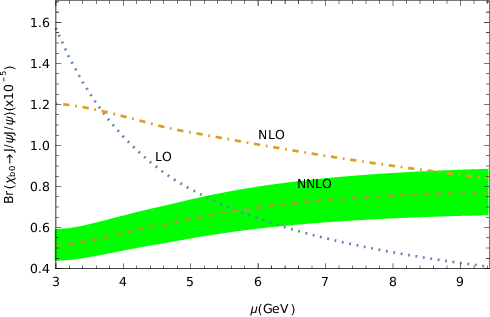}
	\includegraphics*[scale=0.95]{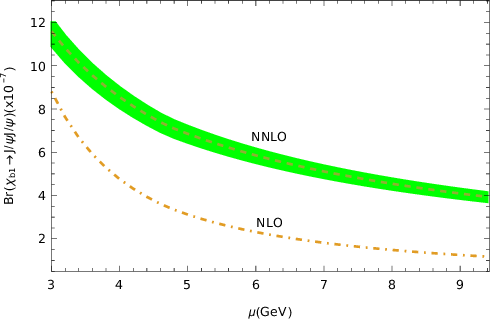}
	\includegraphics*[scale=0.95]{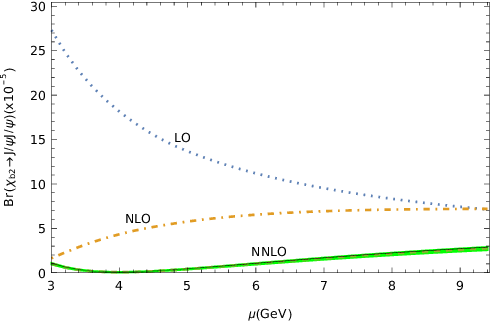}
	\caption{Theoretical predictions for ${\rm Br}[\eta_b(\chi_{bJ})\to J/\psi J/\psi]$ as a function of $\mu_R$ at various levels of accuracy in 
	$\alpha_s$. \label{fig-branching-ratio}}
\end{figure}

In Fig.~\ref{fig-branching-ratio}, we plot the branching fractions as a function of the renormalization scale $\mu_R$ at various level of perturbative accuracy.
The green band corresponds to the uncertainty affiliated with the total decay widths of $\eta_b$ and $\chi_{bJ}$.
We observe that the perturbative corrections seem to considerably reduce the LO $\mu_R$-dependence for $\chi_{b0}$
and slightly reduce the $\mu_R$-dependence for $\chi_{b1}$,  but worsen the $\mu_R$-dependence for $\eta_{b}$.

%----------------------------
%----------------------------

%----------------------------
\begin{figure}[htbp]
	\centering
	\includegraphics*[scale=0.95]{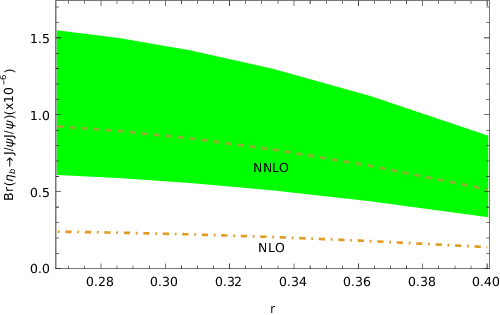}
	\includegraphics*[scale=0.95]{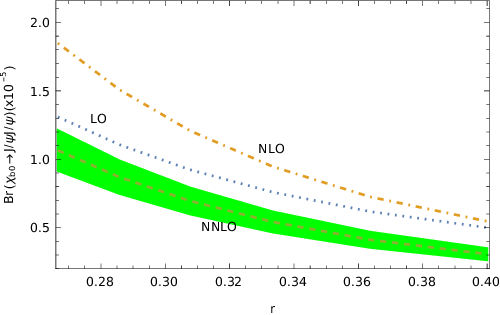}
	\includegraphics*[scale=0.95]{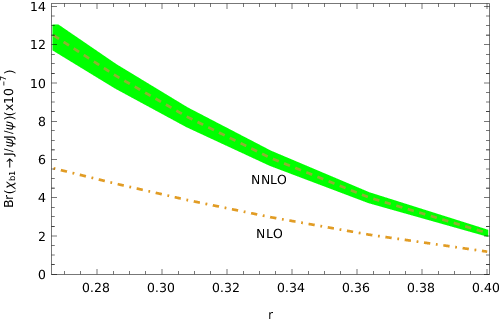}
	\includegraphics*[scale=0.95]{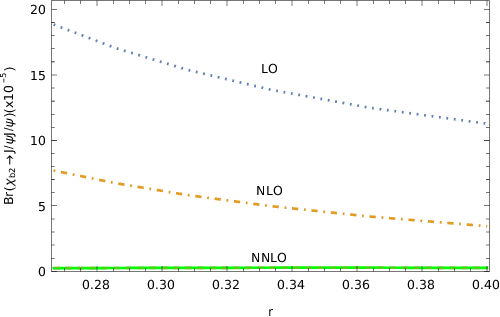}
	\caption{ Branching fractions of $\eta_b(\chi_{bJ})\to J/\psi J/\psi$ as a function of $r$. \label{fig-branching-ratio-mc}}
\end{figure}
%----------------------------

It is instructive to further investigate the dependence of the theoretical predictions on the mass ratio $r$.  
By fixing LDMEs, $\mu_R=m_b$ and $m_b=4.7$ GeV, and varying $m_c$ from $1.25$ GeV to $1.9$ GeV, 
we plot the branching fractions as a function of $r$ at various levels of accuracy in 
$\alpha_s$ in Fig.~\ref{fig-branching-ratio-mc}. 
We observe that the branching fractions monotonically decrease as $r$ increases for $\eta_b$ and $\chi_{b0,1}$ at every perturbative accuracy. 
By analyzing the data in Appendix~\ref{appendix-branching-ratios-mc}, 
we find that, by varying $m_c$ from $1.25$ GeV to $1.9$ GeV, the branching fraction roughly change a factor of $2$, $3$, and $6$ for 
$\eta_b$, $\chi_{b0}$, and $\chi_{b1}$ respectively. 
Although the branching fraction for $\chi_{b2}$ decrease with the increase of $r$  both at LO and NLO,  the branching fraction at NNLO is almost independent 
on $r$.

Finally, we utilize the results in Table~\ref{tab-widths} to estimate the observing prospects at the {\tt LHC} and B factory.
The production cross sections for $\chi_{b0}$ and $\chi_{b2}$ at the {\tt LHC} at $\sqrt{s}=14$ TeV are evaluated: $\sigma(pp \to \chi_{b0} + X) = 1.5\, \mu \rm b$ and
$\sigma(pp \to \chi_{b2} + X) = 2.0\, \mu \rm b$~\cite{Braguta:2005gw}. The cross section for $\eta_b$ production is roughly estimated to be 
$\sigma(pp \to \eta_{b} + X) = 15\, \mu \rm b$~\cite{Jia:2006rx}.
If taking the integrated luminosity $\mathcal{L}=100\, \rm {f b}^{-1}$, it is expected that there will be about $10^6$ exclusive double $J/\psi$ events from $\eta_b$ and $\chi_{b0}$ decay, and $5\times 10^5$ from $\chi_{b2}$ decay at the {\tt LHC}. 
Furthermore, taking into account ${\rm Br}[J/\psi \to  \ell \bar{\ell}] =12\%$, about $(5-10)\times 10^3$ four-lepton events from double $J/\psi$ can be produced.
Although there are potentially copious double $J/\psi$ background events, it might be hopeful to establish the $\eta_b(\chi_{bJ})\to J/\psi J/\psi$ signals.
At the B factory,  $\eta_b$ and $\chi_{bJ}$ can be produced through $\Upsilon(2S)$ electromagnetic E1 transition. Using a sample of $158\times 10^6\, \Upsilon(2S)$
events collected by the {\tt Belle} detector, we expect about $10^5$ $\eta_b$ and $6\times 10^6$ $\chi_{b0}$, and $10^7$ $\chi_{b1,2}$ events can be accumulated.  
Consequently, it is estimated that there are less than $100$ double $J/\psi$ events. So the experimental measurements
on these channels  are challenging  based on the current dataset at the B factory. 
Nevertheless, with the designed $50\, {\rm ab}^{-1}$ integrated luminosity at {\tt Belle} 2, 
it seems that the observation prospects of $\eta_b(\chi_{bJ})\to J/\psi J/\psi$
are promising in the foreseeable future.

%-----------------------------
\section{Summary \label{sec-summary}}
%----------------------------

Based on the NRQCD factorization, 
we compute both the polarized and the unpolarized decay widths for the processes $\eta_b(\chi_{bJ})\to J/\psi J/\psi$ up to NNLO in $\alpha_s$.
By taking the decay width of $\eta_b$ from PDG and determining the total decay widths of $\chi_{bJ}$ through their electromagnetic E1 transition into $\Upsilon$, we also predict the branching fractions for $\eta_b(\chi_{bJ})\to J/\psi J/\psi$. 
We find the perturbative corrections are sizable for $\eta_b$ and $\chi_{b2}$ decay. 
In particular for $\chi_{b2}$,  both the $\mathcal{O}(\alpha_s)$ and $\mathcal{O}(\alpha_s^2)$
corrections can significantly reduce the LO prediction.
Moreover, 
we observe that the decay widths for $\eta_b$ and $\chi_{b1}$ are much smaller than the other two channels, 
which can be attributed to the vanishing LO amplitudes for $\eta_b$ and $\chi_{b1}$.
By including all the radiative corrections, we find the branching fractions are $8.2\times 10^{-7}$, $6.2\times 10^{-6}$, $7.2\times 10^{-7}$, and $2.7\times 10^{-6}$ for $\eta_b$, $\chi_{b0}$,
$\chi_{b1}$, and $\chi_{b2}$ respectively. Our theoretical predictions are consistent with the upper limits measured by the {\tt Belle} Collaboration. 

We also investigate the dependence of the theoretical predictions on
the mass ratio $r$. By fixing $m_b$ and the LDMEs,  and varying $r$ from $0.26$ to $0.4$, we find the 
branching fraction can change a factor of $2$, $3$, and $6$ for $\eta_b$, $\chi_{b0}$, and $\chi_{b1}$, respectively.
Although the  branching fraction for $\chi_{b2}$ decrease with the increase of $r$ at LO and NLO, 
it is almost independent on $r$ at NNLO.

Finally, we explore the observing prospects for $\eta_b(\chi_{bJ})\to J/\psi J/\psi$ at the {\tt LHC} and B factory. 
We expect that there are about $(5-10)\times 10^5$ double $J/\psi$ signal events produced at the {\tt LHC}, and 
less than $100$ events at the B factory.
Taking into account ${\rm Br}[J/\psi \to  \ell \bar{\ell}] =12\%$, about several thousands four-lepton events from double $J/\psi$ can be produced at the {\tt LHC}.
If copious double $J/\psi$ background can be well separated, the experimental measurements on $\eta_b(\chi_{bJ})\to J/\psi J/\psi$ might be hopeful at the {\tt LHC}.
The measurement on $\eta_b(\chi_{bJ})\to J/\psi J/\psi$ at the B factory is quite challenging based on current  dataset. 
Nevertheless, with the designed $50\, {\rm ab}^{-1}$ integrated luminosity at {\tt Belle} 2, 
the observation prospects for $\eta_b(\chi_{bJ})\to J/\psi J/\psi$
may be promising in the foreseeable future.

%---------------------
\section*{Acknowledgments}
%---------------------
The work of Y.-D. Z. is supported by the National Natural Science Foundation
of China under Grants No.~12135006 and No.~12075097,
and by the Fundamental Research Funds for the Central Universities under Grants No.~CCNU20TS007 and No.~CCNU22LJ004.
%---------------------------------
The work of X.-W. B., W.-L. S. and M.-Z. Z. is
supported by the National Natural Science Foundation
of China under Grants No. 12375079 and No. 11975187, and 
the Natural Science Foundation of ChongQing under Grant No. CSTB2023NSCQ-MSX0132.
%---------------------------------
The work of F. F. is supported by the National Natural Science Foundation
of China under Grant No. 12275353.
% and No. 11875318.
%---------------------------------

\section*{Appendix}
\appendix
\section{Construction of helicity projectors\label{appendix-helicity-projectors}}
%------------------------------------------
In this appendix, we present the helicity projectors ${\mathcal P}_{\lambda_1,\lambda_2}^{(H)}$, which have been used to compute the helicity amplitudes for $\eta_b(\chi_{bJ})\to J/\psi J/\psi$ in Sec.~\ref{sec-calculation}. We apply the similar
technique applied in \cite{Xu:2012uh,Zhang:2021ted}.

For convenience, we introduce an auxiliary transverse metric tensor and two auxiliary longitudinal vectors,
%--------------------------
\begin{subequations}\label{eq-auxiliary}
	\begin{eqnarray}
	g_{\perp}^{\mu \nu}&=&g^{\mu \nu}+\frac{P^{\mu} P^{\nu}}{|\mathbf{P}|^{2}}-\frac{Q \cdot P}{m_H^{2}|\mathbf{P}|^{2}}\left(P^{\mu} Q^{\nu}+Q^{ \mu} P^{\nu}\right)+\frac{m^2_{J/\psi}}{m_H^{2}|\mathbf{P}|^{2}}\left(Q^{ \mu} Q^{\nu}\right),\\
	L_H^\mu &=& \frac{1}{|\mathbf{P}|} \bigg(P^\mu-\frac{Q\cdot P}{m_H}Q^\mu\bigg),\\
	L_{J/\psi}^\mu &=& \frac{1}{|\mathbf{P}|} \bigg(\frac{P\cdot Q}{m_H m_{J/\psi}}P^\mu-\frac{m_{J/\psi}}{m_H}Q^\mu\bigg),
	\end{eqnarray}
\end{subequations}
%--------------------------
where $P$ and $Q$ denote the momenta of $J/\psi$ and $H$ respectively.
It is obvious the transverse metric tensor satisfies properties
%---------------------------
\begin{subequations}
\begin{eqnarray}
&&g_{\perp \mu \nu} P^{\mu}=g_{\perp \mu \nu} Q^{\mu}=0,\\
&&g_{\perp \mu}^{\mu}=2,\\
&&g_{\perp \mu\alpha}g_\perp^{\alpha\nu}=g_{\perp \mu\alpha}g^{\alpha\nu}=g_{\perp \mu}^\nu.
\end{eqnarray}
\end{subequations}
The longitudinal vectors satisfy $L_{H}^\mu Q_\mu=L_{J/\psi}^\mu P_\mu=0$.

We enumerate all the eight helicity projectors
%---------------------
\begin{subequations}
	\begin{eqnarray}
	\mathcal{P}_{1,1}^{(\eta_b)\mu \nu}&=&\frac{i}{2 m_{\eta_b}|\mathbf{P}|} \epsilon^{\mu \nu \rho \sigma} Q_{\rho} P_{\sigma},\\
	\mathcal{P}_{1,1}^{(\chi_{b0})\mu\nu}&=&-\frac{1}{2}g_{\perp}^{\mu\nu},\\
	\mathcal{P}_{0,0}^{(\chi_{b0})\mu\nu}&=&L_{J/\psi}^\mu L_{J/\psi}^\nu,\\
	\mathcal{P}_{1,0}^{(\chi_{b1})\mu \nu \alpha}&=&\frac{-1}{2 m_{\chi_{b1}}|\mathbf{P}|} \epsilon^{\mu \nu \rho \sigma} Q_{\rho} P_{\sigma}L_{J/\psi}^\alpha,\\
	\mathcal{P}_{1,-1}^{(\chi_{b2})\mu \nu\alpha\beta}&=&
	\frac{1}{4}\bigg(g_{\perp}^{\mu\nu}g_{\perp}^{\alpha\beta}-g_{\perp}^{\mu\alpha}g_{\perp}^{\nu\beta}
	-g_{\perp}^{\mu\beta}g_{\perp}^{\nu\alpha}\bigg),\\
	\mathcal{P}_{1,1}^{(\chi_{b2})\mu \nu\alpha\beta}&=&
	\frac{-1}{2\sqrt{6}}\bigg(g_{\perp}^{\mu\nu}+2L_{H}^\mu L_{H}^\nu\bigg)g_{\perp}^{\alpha\beta},\\
	\mathcal{P}_{1,0}^{(\chi_{b2})\mu \nu\alpha\beta}&=&
	\frac{-1}{2\sqrt{2}}\bigg(g_{\perp}^{\mu\alpha}L_{H}^\nu+g_{\perp}^{\nu\alpha}L_{H}^\mu\bigg)L_{J/\psi}^\beta,\\
	\mathcal{P}_{0,0}^{(\chi_{b2})\mu \nu\alpha\beta}&=&
	\frac{1}{\sqrt{6}}\bigg(g_{\perp}^{\mu\nu}+2L_{H}^{\mu}L_{H}^{\nu}\bigg)L_{J/\psi}^\alpha L_{J/\psi}^\beta.
	\end{eqnarray}
\end{subequations}

If we express the decay amplitudes of $\eta_b(\chi_{bJ})\to J/\psi(\lambda_1) J/\psi(\lambda_2)$ as
%---------------------
\begin{subequations}
	\begin{eqnarray}
	\mathcal{A}^{(\eta_b)}&=&\mathcal{A}_{\mu \nu}^{(\eta_b)} \epsilon_{ J/\psi}^{*\mu}(\lambda_{1}) \epsilon_{J/\psi}^{*\nu}(\lambda_{2}),\\
	\mathcal{A}^{(\chi_{b0})}&=&\mathcal{A}_{\mu \nu}^{(\chi_{b0})} \epsilon_{ J/\psi}^{*\mu}(\lambda_{1}) \epsilon_{J/\psi}^{*\nu}(\lambda_{2}),\\
	\mathcal{A}^{(\chi_{b1})}&=&\mathcal{A}_{\mu \nu \alpha}^{(\chi_{b1})}\epsilon_{\chi_{b 1}}^{\mu}  \epsilon_{J/\psi}^{*\nu}(\lambda_{1})\epsilon_{J/\psi}^{* \alpha}(\lambda_{2}),\\
	\mathcal{A}^{(\chi_{b2})}&=&\mathcal{A}_{\mu \nu \alpha\beta}^{(\chi_{b2})} \epsilon_{\chi_{b2}}^{\mu \nu} \epsilon_{J/\psi}^{* \alpha}(\lambda_{1})\epsilon_{J/\psi}^{*\beta}(\lambda_2),
	\end{eqnarray}
\end{subequations}
where $\epsilon_{J/\psi}$,$\epsilon_{\chi_{b 1}}$ and $\epsilon_{\chi_{b 2}}$ represent the polarization vector/tensor of the  $J/\psi$, $\chi_{b1}$, and $\chi_{b2}$ respectively,
the helicity amplitude can be computed through
%---------------------
\begin{subequations}
	\begin{eqnarray}
	\mathcal{A}^{(\eta_b)}_{1,1}&=&\mathcal{P}_{1,1}^{(\eta_b)\mu \nu}\mathcal{A}_{\mu \nu}^{(\eta_b)},\\
	\mathcal{A}^{(\chi_{b0})}_{1,1}&=&\mathcal{P}_{1,1}^{(\chi_{b0})\mu \nu}\mathcal{A}_{\mu \nu}^{(\chi_{b0})},\\ 
	\mathcal{A}^{(\chi_{b0})}_{0,0}&=&\mathcal{P}_{0,0}^{(\chi_{b0})\mu \nu}\mathcal{A}_{\mu \nu}^{(\chi_{b0})},\\
	\mathcal{A}^{(\chi_{b1})}_{1,0}&=&\mathcal{P}_{1,0}^{(\chi_{b1})\mu \nu \alpha}\mathcal{A}_{\mu \nu \alpha}^{(\chi_{b1})},\\
	\mathcal{A}^{(\chi_{b2})}_{1,-1}&=&\mathcal{P}_{1,-1}^{(\chi_{b2})\mu \nu\alpha\beta}\mathcal{A}_{\mu \nu\alpha\beta}^{(\chi_{b2})},\\ \mathcal{A}^{(\chi_{b2})}_{1,1}&=&\mathcal{P}_{1,1}^{(\chi_{b2})\mu \nu\alpha\beta}\mathcal{A}_{\mu \nu\alpha\beta}^{(\chi_{b2})},\\
	\mathcal{A}^{(\chi_{b2})}_{1,0}&=&\mathcal{P}_{1,0}^{(\chi_{b2})\mu \nu\alpha\beta}\mathcal{A}_{\mu \nu\alpha\beta}^{(\chi_{b2})},\\ \mathcal{A}^{(\chi_{b2})}_{0,0}&=&\mathcal{P}_{0,0}^{(\chi_{b2})\mu \nu\alpha\beta}\mathcal{A}_{\mu \nu\alpha\beta}^{(\chi_{b2})}.
	\end{eqnarray}
\end{subequations}
%-----------------------------------
\section{Decay widths and branching ratios for different $r$ \label{appendix-branching-ratios-mc}}
%-----------------------------------

By fixing LDMEs, $\mu_R=m_b$ and $m_b=4.7$ GeV, and varying $m_c$ from $1.25$ GeV to $1.9$ GeV, we tabulate 
the decay widths and branching fractions in Table~\ref{tab-mc-dependence}.  The branching fraction as a function of $r$ has been illustrated in Fig.~\ref{fig-branching-ratio-mc}.

\begin{table}[!htbp]\small
	\caption{Theoretical predictions on various unpolarized decay widths (in units of eV) and branching fractions ($\times 10^{-5}$). }
	\label{tab-mc-dependence}
	\centering
	\setlength{\tabcolsep}{2pt}
	\renewcommand{\arraystretch}{1.5}
	%---------------------------------
	%------------------------------------
	\begin{tabular}{|c|c|c|c|c|c|c|c|c|c|c|c|c|c|c|}
		\hline
		\multirow{2}*{H}&$r$&\multicolumn{2}{|c|}{0.267}&\multicolumn{2}{|c|}{0.286}&\multicolumn{2}{|c|}{0.308}&\multicolumn{2}{|c|}{0.333}&\multicolumn{2}{|c|}{0.364}&\multicolumn{2}{|c|}{0.400}\\
		\cline{2-14}
		&Order&$\Gamma$&Br&$\Gamma$&Br&$\Gamma$&Br&$\Gamma$&Br&$\Gamma$&Br&$\Gamma$&Br\\
		\hline
		\multirow{3}*{$\eta_{b}$}
		&LO&--&--&--&--&--&--&--&--&--&--&--&-- \\
		\cline{2-14}
		&NLO&2.407&0.024&2.341&0.023&2.231&0.022&2.059&0.021&1.796&0.018&1.401&0.014 \\
		\cline{2-14}
		&NNLO&9.245&0.092&8.944&0.089&8.468&0.085&7.749&0.077&6.688&0.067&5.151&0.052 \\
		%----------------------------------------
		\hline
		\multirow{3}*{$\chi_{b0}$}
		&LO&14.978&1.309&12.664&1.107&10.563&0.923&8.690&0.760&7.065&0.618&5.722&0.500 \\
		\cline{2-14}
		&NLO&21.144&1.848&17.267&1.509&13.826&1.209&10.831&0.947&8.299&0.725&6.250&0.546 \\
		\cline{2-14}
		&NNLO&12.177&1.064&9.951&0.870&7.956&0.695&6.202&0.542&4.708&0.412&3.502&0.306\\	
		%----------------------------------------
		\hline
		\multirow{3}*{$\chi_{b1}$}
		&LO&--&--&--&--&--&--&--&--&--&--&--&-- \\
		\cline{2-14}
		&NLO&0.044&0.055&0.037&0.047&0.031&0.039&0.024&0.030&0.016&0.021&0.009&0.012 \\
		\cline{2-14}
		&NNLO&0.099&0.125&0.082&0.104&0.065&0.082&0.048&0.061&0.032&0.040&0.017&0.022 \\
		%----------------------------------------
		\hline
		\multirow{3}*{$\chi_{b2}$}
		&LO&33.173&18.848&30.015&17.054&27.096&15.396&24.423&13.877&22.002&12.501&19.848&11.277 \\
		\cline{2-14}
		&NLO&13.545&7.696&11.845&6.730&10.253&5.825&8.764&4.979&7.370&4.187&6.058&3.442 \\
		\cline{2-14}
		&NNLO&0.410&0.233&0.438&0.249&0.459&0.261&0.472&0.268&0.473&0.269&0.58&0.260 \\
		%----------------------------------------
		\hline
	\end{tabular}
\end{table}
%-------------------------------------------------------------------

\end{document}